\documentclass[epjST]{svjour}
\usepackage{graphicx}
\usepackage[utf8]{inputenc}
\usepackage[colorlinks,citecolor=blue,urlcolor=blue,linkcolor=blue]{hyperref}
\journalname{Eur. Phys. J. Special Topics}
\begin{document}
\title{Do supernovae indicate an accelerating universe?}
%
%
\author{Roya Mohayaee\inst{1} 
\and Mohamed Rameez\inst{2} 
\and Subir Sarkar\inst{3}}
\institute{CNRS, UPMC, Institut d'Astrophysique de Paris, 98 bis Bld Arago, Paris, France \and 
Department of High Energy Physics, Tata Institute of Fundamental Research, Homi Bhabha Road, Mumbai 400005, India \and 
Rudolf Peierls Centre for Theoretical Physics, University of Oxford, Parks Road, Oxford OX1 3PU, United Kingdom}
\abstract{
In the late 1990's, observations of two directionally-skewed 
samples of, in total, 93 Type~Ia supernovae were analysed in the framework of the Friedmann-Lema\^{i}tre-Robertson-Walker (FLRW) cosmology. Assuming these to be `standard(isable) candles' it was inferred that the Hubble expansion rate is accelerating as if driven by a positive Cosmological Constant $\Lambda$ in Einstein's theory of gravity. This  is still the only \emph{direct} evidence for the `dark energy' that is the dominant component of today's standard $\Lambda$CDM cosmological model. Other data such as baryon acoustic oscillations (BAO) in the large-scale distribution of galaxies, temperature fluctuations in the cosmic microwave background (CMB), measurement of stellar ages, the rate of growth of structure, \emph{etc} are all `concordant' with this model but do not provide independent evidence for accelerated expansion. The recent discussions about whether the inferred acceleration is real rests on analysis of a larger sample of 740 SNe~Ia which shows that these are not quite standard candles, and more importantly highlights the `corrections' that are applied to analyse the data in the FLRW framework. The latter holds in the reference frame in which the CMB is isotropic, whereas observations are carried out in our heliocentric frame in which the CMB has a large dipole anisotropy. This is assumed to be of kinematic origin i.e. due to our non-Hubble motion driven by local inhomogeneity in the matter distribution which has grown under gravity from primordial density perturbations traced by the CMB fluctuations. The  $\Lambda$CDM model predicts how this peculiar velocity should fall off as the averaging scale is raised and the universe becomes sensibly homogeneous. However observations of the local `bulk flow' are inconsistent with this expectation and convergence to the CMB frame is not seen. Moreover the kinematic interpretation implies a corresponding dipole in the sky distribution of high redshift quasars, which is rejected by observations at $4.9\sigma$. Hence the peculiar velocity corrections employed in supernova cosmology are inconsistent and discontinuous within the data. The acceleration of the Hubble expansion rate is in fact anisotropic at $3.9\sigma$ and aligned with the bulk flow. 
Thus dark energy could be an artefact of analysing data assuming that we are idealised observers in an FLRW universe, when in fact the real universe is inhomogeneous and anisotropic out to distances large enough to impact on cosmological analyses.
} 
\date{Received: date / Accepted: date}
\maketitle
%

\section{Introduction} 
\label{sec:intro}

In the past two decades, a `concordance $\Lambda$CDM model' of cosmology has emerged, in which data are interpreted assuming that the Universe is isotropic and homogeneous, and General Relativity (GR) is the true theory of gravity. In this framework, spacetime is modelled by the maximally-symmetric Friedman-Lema\^{i}tre-Robertson-Walker class of solutions to the Einstein field equations, obtained by requiring the stress-energy tensor to be exactly isotropic and homogeneous on all scales. In the best-fit $\Lambda$CDM model \cite{Zyla:2020zbs} the energy density of the Universe today consists of 4.9\% baryonic matter, 26.8\% dark (gravitating) matter and 68.3\% `dark energy' which has come to dominate the Universe at redshift $z < 1$ and is causing its expansion rate to accelerate --- in its most economical form Einstein's Cosmological Constant $\Lambda$. 

The Cosmological Constant can also be interpreted as the energy density of  fluctuations of the vacuum in the quantum field theories that describe the energy-momentum tensor. Thus the effective $\Lambda$ appearing in the Friedmann-Lemaître equations is the sum of the arbitrary classical contribution (from geometry), and the quantum contribution (from matter) which cannot be rigorously calculated, even its sign. That these unrelated terms should conspire together so as to make $\Lambda$ tiny enough to allow the Universe to have lasted for over 10 billion years --- rather than gone into perpetual inflation or simply recollapsed at a very early age --- is the as yet unsolved Cosmological Constant problem \cite{Weinberg:1988cp}.\footnote{It has been suggested  \cite{Dvali:2018fqu,Dvali:2020etd}, building on previous string-theoretic considerations, that S-matrix formulation of gravity excludes de Sitter vacuua (i.e. $\Lambda$) at the quantum level.}

As is well-documented \cite{Sahni:1999gb,Peebles:2002gy}, $\Lambda$ has had a chequered history with claims for its presence which were subsequently found to be unjustified because of systematic uncertainties (in particular the evolution of `standard candles', `standard rods', galaxy number counts \emph{etc}). Its recent reincarnation was initiated by \emph{indirect} arguments~\cite{Ostriker:1995su} arising of a desire to get diverse cosmological datasets to be concordant with the FLRW model, rather than any compelling physical argument why $\Lambda$ should have a value of ${\cal O}(H_0^2)$, where $H_0 \equiv 100h$~km~s$^{-1}$Mpc$^{-1} \sim 10^{-42}$GeV is the present expansion rate and the only dimensionful parameter in the model.\footnote{This follows from examination of the 
Friedmann-Lemaître equation rewritten as the `cosmic sum rule' for the fractional energy densities in matter, curvature and Cosmological Constant: $\Omega_\textrm{m}+\Omega_{k}+\Omega_{\Lambda}=1$.
Only $\Omega_\textrm{m}$ and $\Omega_{k}$ are directly measured (also the supernova Hubble diagram  determines: $\sim 0.8\Omega_\textrm{m} - 0.6\Omega_{\Lambda}$) and $\Omega_{\Lambda}$ is inferred from the sum rule. Thus any uncertainties in these measurements translate into a non-zero value for $\Omega_{\Lambda}$ \cite{Sarkar:2007cx}. Whether such a $\Lambda$ of ${\cal O}(H_0^2)$ actually exists relies therefore crucially on the validity of this sum rule, i.e. ultimately on the underlying assumptions of \emph{exact} homogeneity and isotropy.} It was not until observations of 93 Type~Ia supernova in the late 1990s claimed \emph{direct} evidence for an accelerating Universe \cite{Perlmutter:1998np,Riess:1998cb}, that $\Lambda$ was taken seriously and $\Lambda$CDM came to be canonised as the `standard model' of cosmology. It was therefore surprising when some years later a principled statistical analysis of a bigger sample of 740 uniformly analysed SNe~Ia in the Joint Lightcurve Analysis (JLA) catalogue \cite{Betoule:2014frx} demonstrated that the evidence for acceleration is rather marginal i.e. $<3\sigma$ \cite{Nielsen:2015pga}. In response it was argued \cite{Rubin:2016iqe} that doing the analysis differently, and in particular taking into account other cosmological data such as on BAO and CMB anisotropies analysed in the FLRW framework, firmly reinstates the acceleration of the universe. To assist the reader in assessing these discussions and to place the history in context, we briefly review  supernova cosmology in Section~\ref{sne}.

 The observed Universe is not quite isotropic. It is also manifestly inhomogeneous. The interpretation of data from the real Universe in terms of the FLRW model has proceeded under the weaker assumptions of \emph{statistical} isotropy and homogeneity, the modern version of the Cosmological Principle. Since the CMB actually exhibits a large dipole anisotropy, this is assumed to be due to our motion with respect to the cosmic rest frame in which the Universe looks FLRW, and data can be analysed according to the Friedman-Lema\^{i}tre equations. Hence cosmological data are corrected for this motion using a special relativistic boost. However as we show in Section~\ref{flrw}, this assumption is no longer tenable. Several independent data sets now argue against the existence of such a frame in the real Universe. At low redshift ($z < 0.1$), all local matter appears to have a coherent `bulk flow' aligned with the direction of the CMB dipole. At higher redshift ($z > 1$) the matter dipole is far larger than is expected from the kinematic interpretation of the CMB dipole.

The peculiar Raychaudhury equation~\cite{Tsagas:2013ila} has been employed to show  \cite{Tsagas:2015mua} that the acceleration of the expansion of space as inferred by an observer embedded within such a bulk flow should show a scale-dependent dipolar modulation. We have tested this hypothesis \cite{Colin:2018ghy}(hereafter: CMRS19) and find that the inferred acceleration of the expansion rate is indeed  directional and its dipole component is 50 times larger than and statistically far more significant (at $3.9 \sigma$) than the isotropic component which is non-zero at only $1.4 \sigma$, suggesting that the inferred acceleration \emph{cannot} be due to $\Lambda$. We also showed (Figure 2 of CMRS19 \cite{Colin:2018ghy}) that SNe~Ia data as disseminated publicly and thus employed by many authors in fitting for concordance cosmology, not only acknowledge the existence of this bulk flow, but `correct' the observations towards an assumed cosmic rest frame in an unrealistic manner using unreliable data sets~\cite{Rameez:2019wdt}. The methodology in CMRS19 \cite{Colin:2018ghy} has been subjected to further  criticism \cite{Rubin:2019ywt} (RH19) which we have rebutted seriatim \cite{Colin:2019ulu}. 

General Relativity (GR), as expressed in the Einstein Field Equations and the Geodesic Equation, is a geometric description of gravity in which spacetime is modelled as a pseudo-Riemannian manifold warped locally by the presence of matter and energy. The real Universe contains structure over a vast range of scales. While any volume element in the FLRW model expands at the same rate as all others, as described by a single scale factor $a$, the expansion of the real Universe in GR is an \emph{average} effect arising from the coarse-graining of small scales and is thus locally different everywhere~\cite{Ellis:2005uz}. This is properly described by the Raychaudhury equation --- which reduces to the Friedmann-Lemaître equation as a limiting case~\cite{Ellis:2007}.\footnote{The freedom to add an arbitrary `Cosmological constant' (multiplied by the metric tensor) to the field equation arises from general covariance~\cite{Norton:1993}, essentially the notion that there are no preferred (either inertial or accelerating) frames in GR. Hence if $\Lambda$ is indeed driving the acceleration then this acceleration must be the same in \emph{all} directions.} Due to these cardinal differences between the real Universe and its simplified description in the FLRW model, even as the `precision cosmology' programme has in the past two decades focused on measuring the $\Lambda$CDM cosmological parameters as precisely as possible, there is an ongoing theoretical debate as to whether this really makes physical sense \cite{Green:2014aga,Buchert:2015iva}. We emphasize that the `fitting problem' in cosmology~\cite{Ellis:1984bqf,Ellis:1987zz} is at the heart of the problem and dedicate Section~\ref{fitting} to a discussion of how this relates to the corrections employed in supernova cosmology. We conclude in Section~\ref{conclusion} that the evidence for acceleration of the Hubble expansion rate from supernovae remains marginal.

\section{The Universe is anisotropic}
\label{flrw}

The `Cosmological Principle' began as an assumption \cite{Peebles:2020} but is now said to be supported by the smoothness of the cosmic microwave background (CMB), which has temperature fluctuations of only about 1 part in 100,000 on small angular scales. These high  multipoles in the CMB angular power spectrum are attributed to Gaussian density fluctuations created in the early universe with a nearly scale-invariant spectrum which have grown through gravitational instability to create the large-scale structure in the present Universe.~\footnote{There are however `low-$\ell$ anomalies' in the CMB anisotrpies, viz. the anomalously small and (planar) aligned quadrupole and octupole, as well as a `hemispherical asymmetry' \cite{Schwarz:2015cma}.} The observed dipole anisotropy of the CMB is however $\sim100$ times larger and is believed to be due to our motion with respect to the `CMB frame' in which it would look isotropic \cite{Sciama:1967zz,Stewart:1967ve}, so is called the kinematic dipole. This motion was attributed to the gravitational effect of the inhomogeneous distribution of matter on local scales, dubbed the ``Great Attractor'' \cite{Dressler:1991}.
In subsequent studies however this `bulk flow' has been seen to be significantly faster and extend deeper than the expectation in the $\Lambda$CDM model \cite{Hudson:2004et,Watkins:2008hf,Carrick:2015xza,Magoulas:2016}. In fact it extends up to and beyond the Shapley supercluster, i.e. out to $> 300$~Mpc \cite{Colin:2010ds,Feindt:2013pma}, which is well beyond the scale of $\sim 100$~Mpc on which the universe is supposedly sensibly homogeneous according to galaxy counts in large-scale surveys \cite{Hogg:2004vw,Scrimgeour:2012wt}. A consistency check would be to measure the concommitant effects on higher multipoles of the CMB anisotropy \cite{Challinor:2002zh}, however the significance of the detection of these is $<3\sigma$ even in the precise measurements by the Planck satellite \cite{Aghanim:2013suk}, which allows up to 40\% of the observed dipole to be due to effects other than our motion \cite{Schwarz:2015cma}.  

Yet another probe is to examine the directional behaviour of the X-ray luminosity-temperature correlation of galaxy clusters which have been mapped out to several hundred Mpc. Examination of the ROSAT catalogue of several hundred clusters too reveals an anisotropy in roughly the same direction at high significance $>5\sigma$ \cite{Migkas:2020fza,Migkas:2021zdo}. 

An independent test would be to check if the reference frame of matter at still greater distances converges to that of the CMB, i.e. the dipole in the distribution of cosmologically distant sources, induced by our motion via special relativistic aberration and Doppler shifting effects, should align both in direction and in amplitude with that inferred from the CMB dipole \cite{Ellis:1984}. However this test when done with the NVSS catalogue of radio sources yielded an anomalously high result \cite{Singal:2011dy}. This was criticised as it was not an all-sky survey, also since the redshifts are not directly measured there is potential for contamination by a `clustering dipole' due to a local inhomogeneity. However the anomaly persists even after such concerns are addressed \cite{Gibelyou:2012ri,Rubart:2013tx,Kothari:2013gya,Colin:2017juj} and is present in many such studies done with different radio catalogues \cite{Bengaly:2017slg,Siewert:2020krp}. Recently the CatWISE catalogue of 1.36 million quasars (with median redshift of 1.2) was analysed in this context and the kinematic expectation for the quasar dipole rejected at $4.9\sigma$ --- the highest significance achieved to date \cite{Secrest:2020has}. Knowing the redshift distribution of the quasars, the expected clustering dipole according to the $\Lambda$CDM model \cite{Gibelyou:2012ri} could be explicitly calculated and shown to be negligible. 

\begin{figure}
\includegraphics[width=1.0\linewidth]{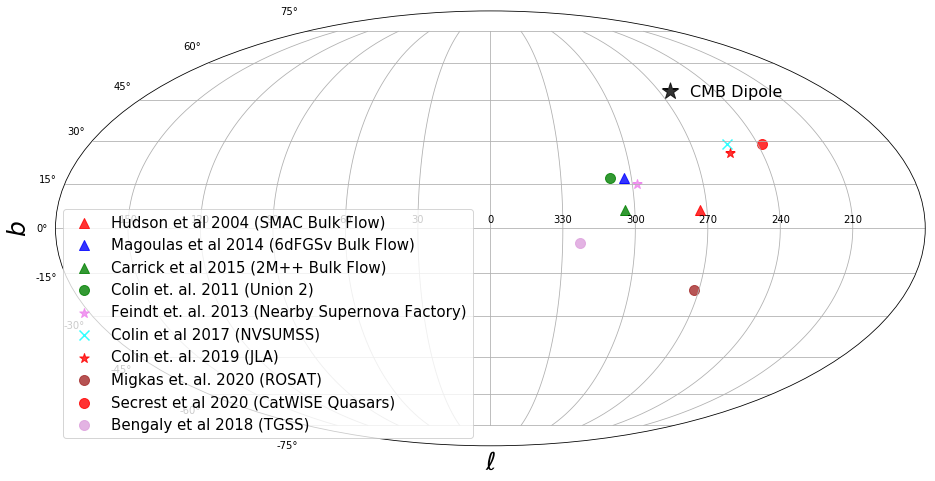}
\caption{Examples of directional anisotropy reported in studies of the local bulk flow \cite{Colin:2010ds,Feindt:2013pma,Hudson:2004et,Carrick:2015xza,Magoulas:2016}, X-ray clusters \cite{Migkas:2020fza,Migkas:2021zdo}, SNe~Ia \cite{Colin:2019ulu}, high redshift radio sources \cite{Colin:2017juj,Bengaly:2017slg} and quasars \cite{Secrest:2020has}. These are all close to the CMB dipole direction \cite{Aghanim:2013suk} which is also marked.}
\label{fig:anisosummary}
\end{figure}

Figure~\ref{fig:anisosummary} shows the directions of the anisotropies reported --- in the local bulk flow traced out to $z \sim 0.1$, in X-ray clusters and SNe~Ia out to $z \sim 1$, and in radio sources and quasars at $z > 1$.

\section{Supernova cosmology}
\label{sne}

Type Ia Supernovae (SNe~Ia) are believed to be thermonuclear explosions in low-mass stars, e.g. triggered when the mass of a white dwarf is driven by the accretion of material from a companion over the maximum that can be supported by electron degeneracy pressure. Since this happens at a critical mass, the Chandrasekhar limit of $\sim 1.4 M_\odot$, all Type~Ia supernovae are taken to have the same intrinsic luminosity, i.e. a `standard candle'. In practice the intrinsic magnitudes of nearby SNe~Ia (to which distances are known via independent means) exhibit a rather large scatter. However by exploiting the observed linear correlation of the (colour-dependent) luminosity decline rate with the peak magnitude~\cite{Phillips:1993ng}, this scatter can be considerably reduced. This makes SNe~Ia `standardisable' candles, i.e. the intrinsic magnitude can be inferred with relatively low scatter (0.1--0.2 mag) by measuring the lightcurves in different (colour) bands~\cite{Leibundgut:2000xw}. Further \emph{assuming} that the intrinsic properties themselves do not evolve with redshift, observations of SNe~Ia can be used to measure the cosmological evolution of the luminosity distance (i.e. of the scale factor) as a function of redshift. 

In detail however the different empirical techniques for implementing the Phillips corrections~\cite{Phillips:1993ng}, viz. the Multi Colour Lightcurve Shape (MLCS) strategy~\cite{Riess:1998cb}, the `stretch factor' corrections \cite{Perlmutter:1998np} and the template fitting or $\Delta m_{15}$ method~\cite{Hamuy:1996ss,Phillips:1999vh}, do \emph{not} agree with each other --- see Figure~4 of Ref.~\cite{Leibundgut:2000xw}. As the sample of SNe~Ia has grown, the tension between the methods has in fact  increased~\cite{Bengochea:2010it}. The MLCS strategy was to simultaneously infer the Phillips corrections and the cosmological parameters using Bayesian inference. However a two-step process --- the `Spectral Adaptive Lightcurve Template' (SALT) --- is now adopted, wherein the shape as well as the colour \cite{Tripp:1997wt} parameters required for the Phillips corrections are first derived from the lightcurve data, and the cosmological parameters are then extracted in a separate step \cite{Guy:2005me}. The current incarnation of this method is SALT2, employed in analysis of recent public SNe~Ia data sets~\cite{Betoule:2014frx,Scolnic:2017caz}, in which every SNe~Ia is assigned three parameters, $m_B^*$, $x_1$ and $c$ --- respectively the apparent magnitude at maximum (in the rest frame `B-band'), the lightcurve shape, and the lightcurve colour correction. This can be used to construct the distance modulus using the Tripp formula \cite{Tripp:1997wt}:
\begin{equation}
\label{eq:distmod}   
\mu_\mathrm{SN} = m_B^* - M^0_B + \alpha x_1 - \beta c ,
\end{equation}
where $M^0_B$ is the absolute magnitude (degenerate with the absolute distance scale i.e. the value of $H_0$) while $\alpha$ and $\beta$ are parameters which are assumed to be constants for all SNe~Ia. (Further parameters can be added, e.g. a `mass step correction' according to the mass of the SNe~Ia host galaxy, but this turns out to be irrelevant in the fitting exercise, whereas the stretch and colour corrections above are both important and uncorrelated with each other \cite{Nielsen:2015pga}.)  This is related to the luminosity distance $d_\mathrm{L}$ as
\begin{equation}
\mu = 25 + 5 \mathrm{log}_{10} (d_\mathrm{L}/\mathrm{Mpc}),
\label{eq:distmodm}
\end{equation}
where $d_\mathrm{L}$ is a function of the $\Lambda$CDM model parameters:

\begin{eqnarray}
& d_\mathrm{L} &= (1 + z) \frac{d_\mathrm{H}}{\sqrt{\Omega_k}} 
 \mathrm{sin}\left(\sqrt{\Omega_k} \int_0^{z} \frac{H_0 \mathrm{d}z'}{H(z')}\right), \mathrm{for \ } \Omega_k > 0 \\
&  &= (1 + z) d_\mathrm{H} \int_0^{z} \frac{H_0 \mathrm{d}z'}{H(z')}, \mathrm{for \ } \Omega_k = 0 \nonumber \\
&  &= (1 + z) \frac{d_\mathrm{H}}{\sqrt{\Omega_k}} 
 \mathrm{sinh}\left(\sqrt{\Omega_k} \int_0^{z} \frac{H_0 \mathrm{d}z'}{H(z')}\right), \mathrm{for \ } \Omega_k < 0 \nonumber \\
\mathrm{where:}& d_\mathrm{H} &= c/H_0, \quad H_0 \equiv 
 100h~\mathrm{km}\,\mathrm{s}^{-1}\mathrm{Mpc}^{-1}, \nonumber \\ 
& H &= H_0 \sqrt{\Omega_\mathrm{m} (1 + z)^3 + \Omega_k (1 + z)^2  + \Omega_\Lambda} \nonumber.
\label{eq:DLEQ}
\end{eqnarray}
Here $H$ the Hubble parameter ($H_0$ being its present value), $d_\mathrm{H}$ is called the `Hubble distance', and $\Omega_\mathrm{m} \equiv \rho_\mathrm{m}/(3H_0^2/8\pi G_\mathrm{N}), \Omega_\Lambda \equiv \Lambda/3H_0^2, \Omega_k \equiv -kc^2/H^2_0 a^2_0$ are the matter, cosmological constant and curvature densities in units of the critical density which are related in the $\Lambda$CDM model by the `cosmic sum rule': $1=\Omega_\mathrm{m} + \Omega_\Lambda + \Omega_k$.  

To extract cosmological parameters, this distance modulus needs to be compared to the theoretical prediction from a model, using an appropriate and principled statistical procedure.

\begin{table}
\centering
\resizebox{\textwidth}{!}{%
\begin{tabular}{|l|c|c|c|c|c|c|}
\hline
Collaboration &Number of SNe~Ia &$N_\mathrm{out}$ &Lightcurve &CRF &Treatment of peculiar velocities &Lensing\\
\hline
SCP \cite{Perlmutter:1998np} &60 (18+42) &4 &``stretch" &LG &$\sigma_v = 300$~km~s$^{-1}$ &\\
HZT \cite{Riess:1998cb} &50 (34+16) &- &MLCS, template &CMB &$\sigma_v = 200$~km~s$^{-1}$, 2500~km~s$^{-1}$ at high $z$&\\
SNLS \cite{Astier:2005qq} &117 (44+73) &2$^a$ &SALT &CMB+helio &$z_\mathrm{min}=0.015$&\\
SCP (Union) \cite{Kowalski:2008ez} &307&8$^a$ &SALT &CMB+helio &$z_\mathrm{min}=0.015$, $\sigma_v = 300$~km~s$^{-1}$  &$\sigma^l=0.093z$\\
Union2 \cite{Amanullah:2010vv} &557 &12$^a$ &SALT2 &CMB+helio &$z_\mathrm{min}=0.015$, $\sigma_v = 300$~km~s$^{-1}$ &$\sigma^l=0.093z$\\
SCP \cite{Suzuki:2011hu} &580 &0 &SALT2 &CMB+helio &not available &corrections\\
SNLS \cite{Conley:2011ku} &472 &6$^a$ &SALT2 \& SiFTO &CMB &$\sigma_v = 150$~km~s$^{-1}$ + SN-by-SN corrections &$\sigma^l=0.055z$\\
JLA \cite{Betoule:2014frx} &740 &0 &SALT2 &CMB &$\sigma_v = 150$~km~s$^{-1}$ + SN-by-SN corrections &$\sigma^l=0.055z$\\
Pantheon \cite{Scolnic:2017caz} &1048 &86 &SALT2 &CMB &$\sigma_v = 250$~km~s$^{-1}$ + SN-by-SN corrections &$\sigma^l=0.055z$\\
\hline
\end{tabular}
}
\caption{Summary of approaches in major SNe~Ia cosmology analyses. $N_\mathrm{out}$ is the number of outliers rejected, while CRF is the choice made of Cosmic Rest Frame. The first analyses argued that gravitational lensing is a negligible effect, however subsequently a $z$-dependent systematic uncertainty to account for it has been introduced.\\
$^a$ Explicitly noted as ``3 $\sigma$ outlier'' rejections.}
\label{tab:SNeIasummary}
\end{table}

\subsection{Standardisable candles?}

Using SNe~Ia to measure cosmological parameters involves the implicit assumption that their intrinsic properties do \emph{not} evolve with redshift. This requires that the SN luminosity after standardisation ($ m_B^* + \alpha x_1 -\beta c$) should be invariant or equivalently that $M^0_B$ should not evolve with look back time. This assumption has been challenged~\cite{Kang:2019azh,Lee:2020usn}, although this too is disputed~\cite{Rose:2020shp}.
In addition, it has been argued that the distributions of $x_1$ and $c$ are explicitly sample- and redshift-dependent \cite{Rubin:2016iqe,Rubin:2019ywt} even though this requires adding 12 new parameters to a 10-parameter model, so is not justified by any information criterion \cite{Colin:2019ulu}. If that is indeed the case, it is unclear how SNe~Ia can still be thought of as standardisable candles. This would require the other terms in Eq.(\ref{eq:distmod}) to evolve with redshift at just the right rate such that $M^0_B$ remains invariant. If the absolute magnitude were to evolve with redshift this would of course trivially undermine the inference of accelerated expansion \cite{Tutusaus:2017ibk,Tutusaus:2018ulu}.

Since the discussion on this issue~\cite{Kang:2019azh,Rose:2020shp} hinges on how the intrinsic scatter (of unknown origin) of SNe~Ia data in the Hubble diagram is treated statistically, a closer examination of the usual method is in order.

\subsection{The statistical method}

Historically, analyses in supernova cosmology  \cite{Astier:2005qq,Kowalski:2008ez,Amanullah:2010vv,Conley:2011ku,Suzuki:2011hu} have relied on a `constrained $\chi^2$ statistic' defined as follows: 
\begin{equation}
\label{eq:chisq}   
\chi^2 = \sum_i \frac{(\mu_i - \mu_i^\mathrm{obs})^2}{\sigma_{\mu i}^2} .
\end{equation}
Here $\mu_i^\mathrm{obs}$ is the distance modulus for the $i^{th}$ SNe~Ia derived from observations by employing the Phillips corrections according to Eq.(\ref{eq:distmod}) and $\mu_i$ is the corresponding model prediction based on its redshift (Eq.\ref{eq:distmodm}). The variance $\sigma_{\mu i}^2$ is the sum of different sources of uncertainty added in quadrature. 

\begin{equation}
\label{eq:sigmamu}   
\sigma_{\mu i}^2 = (\sigma_{\mu i}^\mathrm{fit})^2 + (\sigma_{\mu i}^{z})^2 + (\sigma_{\mu,i}^l)^2 +  (\sigma_{\mu}^\mathrm{int})^2 ,
\end{equation}
where $\sigma_{\mu i}^\mathrm{fit}$ is the uncertainty from the lightcurve fitting process,  $\sigma_{\mu i}^{z}$ is the uncertainty of the SN redshift measurement, from spectroscopy as well the peculiar velocities of, and within, the host galaxy, and $\sigma_{\mu,i}^l$ is the uncertainty due to gravitational lensing. Since each SNe~Ia observation corresponds to a datum that is still intrinsically scattered around the expectation from the fiducial cosmological model, an additional term $\sigma_{\mu}^\mathrm{int}$ is introduced and estimated from the fit by \emph{requiring} that $\chi^2/\mathrm{d.o.f.} = 1$, sometimes on a sample-by-sample basis. This is very \emph{ad hoc} and statistically unprincipled~\cite{Gull:1989,Karpenka:2015vva}. Since $\chi^2/\mathrm{d.o.f.}$ is the statistic that is usually used to judge goodness of fit, requiring it to be 1 (i.e. a perfect fit) as a condition to estimate yet another parameter ($\sigma_{\mu}^\mathrm{int}$) forsakes any ability to judge the  goodness-of-fit of the assumed model, and no confidence interval can be provided on the $\sigma_{\mu}^\mathrm{int}$ thus estimated. In essence a large effective number of trials associated with trying various values of $\sigma_{\mu}^\mathrm{int}$ in the process of arriving at the final value have been ignored. 

It is motivated by these considerations that the Maximum Likelihood Estimator method ~\cite{Nielsen:2015pga,March:2011xa} was introduced for supernova cosmology. Ref. \cite{Karpenka:2015vva} notes that if $x_1$ and $c$ evolve with redshift, the likelihood-based methods return biased values of the parameters (while the `constrained $\chi^2$' method continues to be robust), however this conclusion is arrived at using Monte Carlo simulations which \emph{assume} the $\Lambda$CDM model and is therefore a circular argument. It has been emphasised \cite{Dam:2017xqs} that systematic uncertainties and selection biases in the data need to be corrected for in a model-independent manner, \emph{before} fitting to a particular cosmological model.

An examination of previous SNe~Ia analyses (see Table~\ref{tab:SNeIasummary}) reveals that the choice of the component of $\sigma_{\mu i}^{z}$ from peculiar velocities as well as $\sigma_{\mu i}^{l}$ the uncertainty due to lensing, has varied considerably. As is shown in figure B.1. of CMRS19, while $\sigma_{\mu i}^\mathrm{int}$ is independent of redshift, the peculiar velocity component of $\sigma_{\mu i}^{z}$ varies inversely with the redshift of the SNe~Ia, while $\sigma_{\mu i}^{l}$ is proportional to the redshift.

\subsection{The accelerated expansion of the Universe}

\begin{figure}
\begin{center}
\includegraphics[width=0.49\linewidth]{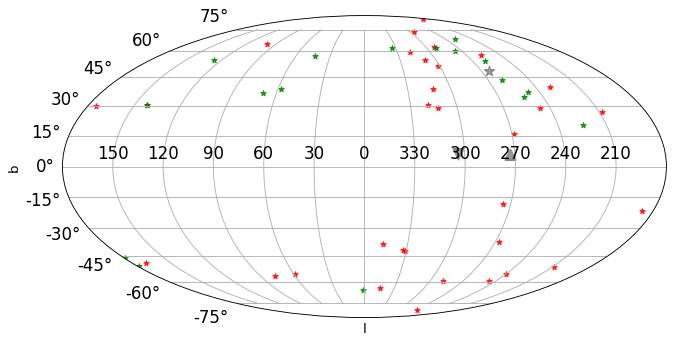}
\includegraphics[width=0.49\linewidth]{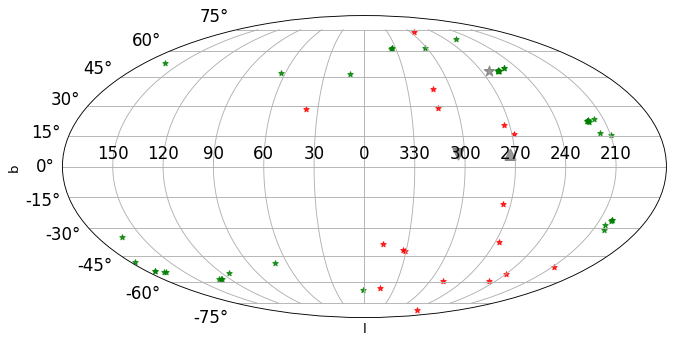}
\caption{Sky distribution of the SNe~Ia used by the HZT \cite{Riess:1998cb} (left) and the SCP \cite{Perlmutter:1998np} (right). Supernovae at $z < 0.1$ are indicated in red while those at $z > 0.1$ are in green.}
\label{fig:riessandperlskymap}
\end{center}
\end{figure}

The original analyses by the Supernova Cosmology Project (SCP)~\cite{Perlmutter:1998np} and High z Supernova Search Team (HZT)~\cite{Riess:1998cb} employed, respectively, 60 and 50 SNe~Ia, 17 of which were in common. Subsequent tests on larger datasets have shown that the evidence is strongly dependent on the assumption of isotropy \cite{Seikel:2008ms}, while supernova datasets themselves show evidence for a local bulk flow, i.e. \emph{anisotropy} \cite{Colin:2010ds,Feindt:2013pma}. CMRS19 \cite{Colin:2018ghy} found that the inferred acceleration is also highest in this general direction when the analysis is done in the heliocentric frame. In Figure~\ref{fig:riessandperlskymap} we show that most of the SNe~Ia observed by SCP and HZT were in fact also in the same general direction.

Nielsen et al.\cite{Nielsen:2015pga} showed using a Maximum likelihood estimator that even with the 740 SNe~Ia in the much enlarged JLA catalogue \cite{Betoule:2014frx}, uniformly analysed with the SALT2 template, the evidence for acceleration is quite marginal ($\sim2.8\sigma$). This analysis employed the JLA catalogue as it ships, i.e. with peculiar velocity (PV) corrections. Since such corrections introduce an arbitrary discontinuity within the data and modify the low redshift SNe~Ia which serve to fix the lever arm of the Hubble diagram~\cite{Colin:2018ghy}, the impact of these corrections on the evidence for cosmic acceleration is worth examining. (Note that PV corrections in SNe~Ia analyses were first introduced only in 2011 \cite{Conley:2011ku}.)

Over half of the 740 SNe~Ia in the JLA catalogue are relatively local ($z < 0.1$) and the furthest is at $z = 1.4$. Hence the luminosity distance $d_\mathrm{L}$ can be quite accurately, to within 7\% of Eq.(\ref{eq:DLEQ}), expanded as a Taylor series in terms of the Hubble parameter $H_0$, the deceleration parameter $q_0 \equiv -\ddot{a} a/\dot{a}^2$, and the jerk $j_0 \equiv \dot{\ddot{a}}/a H^3$. This is `cosmography', i.e. independent of assumptions about the content of the universe. (Specifically in the $\Lambda$CDM model, $q_0 \equiv \Omega_\mathrm{m}/2 - \Omega_\Lambda$.) Modified to explicitly show the dependence on the measured heliocentric redshifts, this writes (RH19):
\begin{equation}
\label{eq:Qkin}    
d_\mathrm{L} (z, z_\mathrm{hel}) = \frac{cz}{H_0} \left[1 + \frac{1}{2}(1 - q_0)z - \frac{1}{6} (1-q_0 - 3q^2_0 + j_0 - \Omega_k) z^2 \right] \times \frac{1+z_\mathrm{hel}}{1+z} ,
\end{equation}
where $z$ can be the measured heliocentric redshift, boosted to the CMB frame, or boosted to the CMB frame with further PV `corrections' applied.

\begin{figure}
\begin{center}
\includegraphics[width=0.49\linewidth]{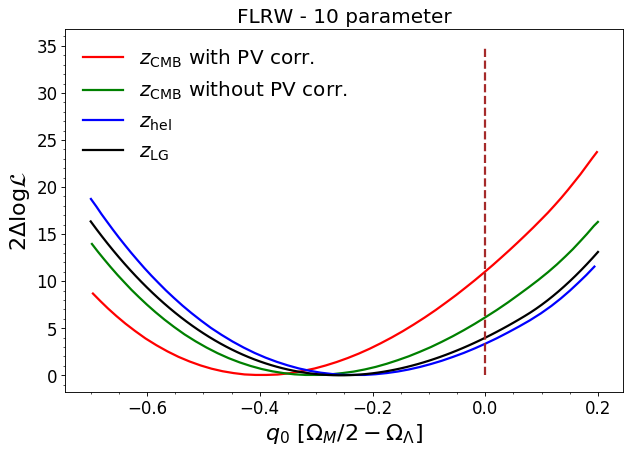}
\includegraphics[width=0.49\linewidth]{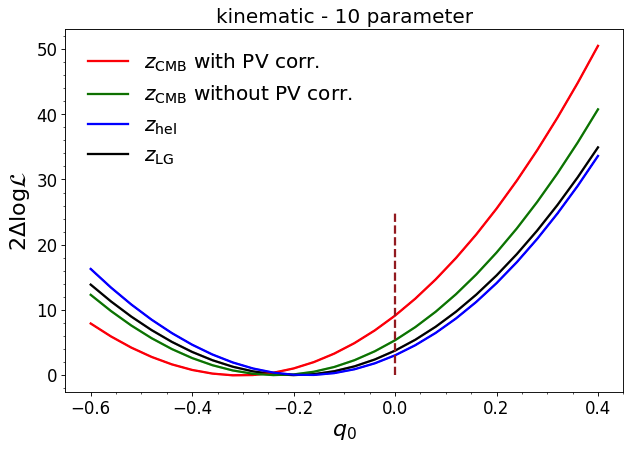}
\caption{Left: The profile likelihood for the 10-parameter FLRW analysis (\ref{eq:DLEQ}), using the JLA~\cite{Betoule:2014frx} colour $c$ and stretch $x_1$ corrections --- see Table~\ref{tab:llhfitparsFLRW}. Right: The same for the cosmographic analysis (\ref{eq:Qkin}) --- see Table~\ref{tab:llhfitparskinem}.}
\label{fig:proflikescan}
\end{center}
\end{figure}

\begin{figure}
\begin{center}\includegraphics[width=0.49\linewidth]{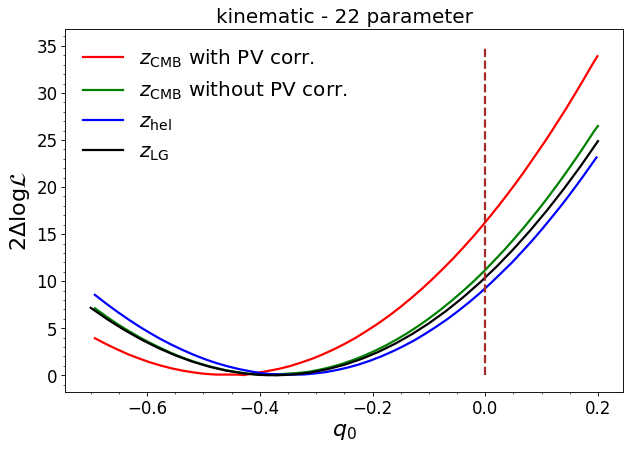}
\caption{The profile likelihood for the 22-parameter cosmographic analysis of RH19 \cite{Rubin:2019ywt}, employing the sample- and redshift-dependent treatment of $c_0$ and $x_{1,0}$ \cite{Rubin:2016iqe} --- see  Table~\ref{tab:llhfitparskinemRH2}.}
\label{fig:proflikescan22pars}
\end{center}
\end{figure}

We redo the MLE analysis \cite{Nielsen:2015pga} in four different ways and present the results in Figures~\ref{fig:proflikescan} and \ref{fig:proflikescan22pars}, and in Tables~\ref{tab:llhfitparsFLRW} and \ref{tab:llhfitparskinem}, respectively --- both for the standard $\Lambda$CDM model (\ref{eq:DLEQ}) and the cosmographic Taylor expansion (\ref{eq:Qkin}). For each case we also show the fit quality when $q_0$ is held at zero (``No accn.").


\begin{enumerate}

\item $z_\mathrm{CMB}$ with PV corr.: This reproduces the earlier analysis of Ref.\cite{Nielsen:2015pga}. The CMB frame redshifts are used, with further corrections made for the peculiar velocities of the SNe~Ia w.r.t. the CMB frame, and the peculiar velocity covariance matrix is included.

\item $z_\mathrm{CMB}$ without PV corr.:  Now CMB frame redshifts are used \emph{without} correcting for the flow of the SNe~Ia w.r.t. this frame and the peculiar velocity component of the covariance matrix is excluded. Note that transforming from heliocentric to CMB frame redshifts requires assuming that the CMB dipole is kinematic in origin. 

\item $z_\mathrm{hel}$: Heliocentric redshifts are used, no corrections are employed and the peculiar velocity component of the covariance matrix is excluded. 

\item $z_\mathrm{LG}$: Finally, redshifts in the Local Group frame are used and the peculiar velocity component of the covariance matrix is excluded. This is similar to what was done by the SCP \cite{Perlmutter:1998np}. 

\end{enumerate}
Our results in Tables~\ref{tab:llhfitparsFLRW} and \ref{tab:llhfitparskinem} illustrate that the peculiar velocity corrections serve to bias the data towards \emph{higher} acceleration (more negative $q_0$). Using observables corrected to the Local Group, as employed by SCP \cite{Perlmutter:1998np}, the change in $2\mathrm{log}\mathcal{L}$ between the best-fit model and the one with zero acceleration is only 4.0, indicating that the preference for acceleration is $<1.6\sigma$. 

\begin{table*}
\resizebox{\textwidth}{!}{%
\begin{tabular}{| l | c | c | c | c | c | c | c | c | c | c | c | c|}
\hline
 Fit & -2log$\mathcal{L}_\mathrm{max}$ & $\Omega_\mathrm{m}$ & $\Omega_{\Lambda}$ & $\alpha$ & $x_{1,0}$ & $\sigma_{x_{1,0}}$ & $\beta$ & $c_0$  & $\sigma_{c_0}$ & $M_0$ & $\sigma_{M_0}$ \\ 
\hline\hline
1. $z_\mathrm{CMB}$ w. PV corr.  &-221.9&0.3402 &0.5653&0.1334&0.03849&0.9321&3.056&-0.01584&0.07107&-19.05&0.1073 \\
As above + No accn. & -211.0 &0.0699&0.03495&0.1313&0.03275&0.9322&3.042&-0.01318&0.07104&-19.01&0.1087 \\
\hline\hline
2. $z_\mathrm{CMB}$ w/o PV corr. &-210.5&0.2828&0.4452&0.135&0.0389&0.9315&3.024&-0.01686&0.07109&-19.04&0.11 \\
As above + No accn. &-204.4&0.07345&0.03673&0.1334&0.03451&0.9316&3.013&-0.01491&0.07105&-19&0.1109\\
\hline\hline
3. $z_\mathrm{hel}$, w/o PV corr.&-198.2&0.2184&0.3387&0.1333&0.03974&0.9317&3.018&-0.01448&0.07111&-19.03&0.1116\\
As above + No accn. &-194.9&0.05899&0.0295&0.1322&0.03649&0.9317&3.006&-0.01302&0.07108&-19&0.1122 \\
\hline\hline
3. $z_\mathrm{LG}$, w/o PV corr.&-193.3&0.2283&0.3708&0.1323&0.0399&0.9317&2.999&-0.01327&0.07114&-19.04&0.1177\\
As above + No accn. &-189.3&0.05083&0.02541&0.131&0.03635&0.9317&2.986&-0.01168&0.07111&-19.01&0.1184
 \\
\hline
\end{tabular}
}
\caption{Results of a 10-parameter fit to the $\Lambda$CDM model (\ref{eq:DLEQ}), using the JLA~\cite{Betoule:2014frx} colour ($c$) and stretch ($x_1$) corrections.}
\label{tab:llhfitparsFLRW}
\end{table*}

\begin{table}
\centering
\resizebox{\textwidth}{!}{%
\begin{tabular}{| l | c | c | c | c | c | c | c | c | c | c | c | c|}
\hline
 Fit & -2log$\mathcal{L}_\mathrm{max}$ & $q_0$ & $j_0 - \Omega_k$ & $\alpha$ & $x_{1,0}$ & $\sigma_{x_{1,0}}$ & $\beta$ & $c_0$  & $\sigma_{c_0}$ & $M_0$ & $\sigma_{M_0}$ \\ 
\hline\hline
1. $z_\mathrm{CMB}$ w. PV corr. &-220.8&-0.311&0.02946&0.1332&0.03809&0.9323&3.056&-0.01591&0.07108&-19.05&0.1074 \\
 As above + No accn. & -211.4&0&-0.8211&0.1312&0.03267&0.9321&3.041&-0.01295&0.07102&-19.01&0.1087 \\
\hline\hline
2. $z_\mathrm{CMB}$ w/o PV corr. &-210.1&-0.2332&-0.2328&0.1348&0.03859&0.9317&3.023&-0.01689&0.07109&-19.03&0.11\\
 As above + No accn. &-204.8&0&-0.8183&0.1333&0.03446&0.9315&3.012&-0.01473&0.07104&-19&0.1108\\
\hline\hline
3. $z_\mathrm{hel}$, w/o PV corr. &-198.1 &-0.1764&-0.4405&0.1332&0.03955&0.9318&3.017&-0.01449&0.07111&-19.03&0.1116 \\
As above + No accn. &-195.1 &0&-0.8534&0.1321&0.03645&0.9317&3.006&-0.01288&0.07106&-19&0.1122 \\
\hline\hline
3. $z_\mathrm{LG}$, w/o PV corr. &-193.3&-0.2005&-0.3856&0.1322&0.0397&0.9318&2.998&-0.01329&0.07114&-19.04&0.1177 \\
As above + No accn. &-189.6&0&-0.8685&0.131&0.03629&0.9317&2.985&-0.01151&0.0711&-19.01&0.1184\\
\hline
\end{tabular}
}
\caption{Results of a 10-parameter fit in the cosmography framework (\ref{eq:Qkin}), using the JLA~\cite{Betoule:2014frx} stretch ($x_1$) and colour ($c$) corrections.}
\label{tab:llhfitparskinem}
\end{table}

In Table~\ref{tab:llhfitparskinemRH2} we adopt the redshift-dependent treatment of the stretch and colour corrections advocated by RH19 \cite{Rubin:2016iqe,Rubin:2019ywt} ($x_{1,0}$ and $c_0$ being their present values) and use redshifts transformed to the CMB frame. Nevertheless the evidence for acceleration remains $<3\sigma$ unless further `corrections' are made for the peculiar velocities of the SNe~Ia with respect to the CMB frame. As shown by CMRS19 \cite{Colin:2018ghy}, these corrections asssume that convergence to the CMB frame is achieved by $\sim150$~Mpc even though the data say otherwise \cite{Colin:2010ds,Feindt:2013pma,Hudson:2004et,Carrick:2015xza,Magoulas:2016}.

\begin{table*}
\centering
\resizebox{\textwidth}{!}{%
\begin{tabular}{| l | c | c | c | c | c | c | c |}
\hline
Fit & -2log$\mathcal{L}_\mathrm{max}$ & $q_0$ & $j_0 - \Omega_k$  & $\alpha$ &  $\beta$ & $M_0$ & $\sigma_{M_0}$ \\ 
\hline\hline
1. $z_\mathrm{CMB}$ w. PV corr.& -339.5&-0.4577&0.1494&0.1334&3.065&-19.07&0.1065 \\
As above + No accn. &-323.3&0&-1.349&0.1312&3.046&-19.01&0.108 \\
\hline\hline
2. $z_\mathrm{CMB}$ w/o PV corr. &-328.3&-0.3776&-0.1781&0.1349&3.034&-19.06&0.1091 \\
As above + No accn. &-317.2&0&-1.326&0.1331&3.017&-19.01&0.1102\\
\hline\hline
3. $z_\mathrm{hel}$ w/o PV corr. &-316.1&-0.3448&-0.3651&0.1333&3.027&-19.05&0.1105\\
As above +  No accn. &-306.9&0&-1.378&0.1317&3.005&-19.01&0.1117 \\
\hline\hline
4. $z_\mathrm{LG}$ w/o PV corr. &-311.7&-0.3750&-0.2809&0.1323&3.007&-19.06&0.1165\\
As above +  No accn. &-301.3&0&-1.413&0.1323&2.982&-19.02&0.1178 \\
\hline
\end{tabular}
}
\caption{Selected results from the 22-parameter fit in the cosmography framework (\ref{eq:Qkin}) using sample- and redshift-dependent stretch ($x_{1}$) and colour ($c$) corrections \cite{Rubin:2016iqe,Rubin:2019ywt}.}
\label{tab:llhfitparskinemRH2}
\end{table*}

It is unrealistic to make such corrections for peculiar velocities which leave the SNe~Ia  immediately outside the flow volume uncorrected \cite{Colin:2018ghy}, in particular such a procedure induces a directional bias (which in the JLA uncertainty budget is simply assigned an uncorrelated variance of $c\sigma_z = 150$~km~$s^{-1}$). An isotropic acceleration can be extracted from the data only by `correcting' over half of all observed supernovae to the CMB frame --- to which convergence has never been demonstrated. 

In fact the peculiar velocity corrections affect  the lever arm of the Hubble diagram in a non-obvious manner. The subsequent Pantheon compilation \cite{Scolnic:2017caz} initially included peculiar velocity corrections far beyond the extent of any actual survey or even flow model.\footnote{\href{https://github.com/dscolnic/Pantheon/issues/2}{https://github.com/dscolnic/Pantheon/issues/2}.} When this issue was pointed out and the bug was fixed, both the magnitudes and heliocentric redshifts of the corresponding supernovae were noted to be discrepant \cite{Rameez:2019wdt}.\footnote{\href{https://github.com/dscolnic/Pantheon/issues/3}{https://github.com/dscolnic/Pantheon/issues/3}.} We cannot analyse data from subsequent surveys such as the Carnegie Supernova Project \cite{Krisciunas:2017yoe} and the Dark Energy Survey \cite{Abbott:2018wog} since neither dataset is publicly available in an usable form, unlike the JLA catalogue which provides full details of the lightcurve fitting parameters and all individual covariances.

\section{Peculiar velocity corrections \& the `fitting problem'}
\label{fitting}

We can observe only one Universe from our unique vantage point. As pointed out in Ref.\cite{Ellis:1987zz} any anisotropies we observe are direct measures of the non-FLRW nature of the real Universe. Fitting an idealised FLRW model to the real Universe necessarily involves a choice of `corresponding 2-spheres' which then fixes the cosmic rest frame (CRF) in the real Universe with respect to which all non-Hubble velocities are defined.  

The cosmological redshift ($z_\mathrm{c}$) in the cosmic rest frame is modified by peculiar velocities of the source and the observer as \cite{Ellis:1987zz}:

\begin{equation}
\label{eq:zpv}   
1+z = (1 + z_\mathrm{O})(1 + z_\mathrm{c})(1 + z_\mathrm{s})
\end{equation}
\noindent 
where $z$ is the measured redshift in some reference frame  and $z_\mathrm{O}$ ($z_\mathrm{s}$) is the redshift due to the peculiar velocity of the observer (source) with respect to this frame. (Until 2008, the approximate expression $z \sim z_\mathrm{c} + z_\mathrm{O} + \ldots$ was used instead of the correct one above~\cite{Davis:2010jq}.) Ref.\cite{Ellis:1987zz} emphasised that in addition to  a best-fit FLRW model given available data, a statement of the goodness-of-fit is also necessary to determine the extent of the non-FLRW nature of the real Universe. The `constrained $\chi^2$' method is however by construction unsuited to judge goodness-of-fit, while the choice of CRF has also evolved (see Table~\ref{tab:SNeIasummary}). Thus the procedures adopted in supernova cosmology are at odds with the fundamental challenge of the `fitting problem' in cosmology.

\section{Conclusions}
\label{conclusion}

We have shown \cite{Colin:2018ghy} that the  acceleration of the Hubble expansion rate inferred from SNe~Ia magnitudes and redshifts as measured (in the heliocentric frame) is described by a dipole anisotropy. To infer an isotropic component of the acceleration i.e. a monopole (such as can be attributed to $\Lambda$), it is necessary to boost to the (possibly mythical) CMB frame, and `correct' the redshifts of the low-$z$ supernovae further for their motion w.r.t. the CMB frame \cite{Rubin:2019ywt}. This is done using bulk flow models \cite{Hudson:2004et,Carrick:2015xza} which already  \emph{assume} that the universe is well-described by the standard $\Lambda$CDM model. Moreover to boost the significance of the (monopole) acceleration above $\sim3\sigma$ the supernova lightcurve parameters need to be empirically modelled \emph{a posteriori} as being both sample- and redshift-dependent \cite{Rubin:2016iqe} --- violating both basic principles of unbiased hypothesis testing and the Bayesian Information Criterion ~\cite{Gao:2017}. Only by doing so can the significance of acceleration be raised to $4.2 \sigma$ \cite{Rubin:2016iqe,Rubin:2019ywt}.

It is often said that there is independent evidence for acceleration from observations of CMB anisotropies and large-scale structure. This is \emph{not} the case, in particular all low-redshift probes such as BAO, cosmic chronometers $H(z)$ and the growth rate of structure $\sigma_8(z)$, are also consistent with a non-accelerating universe \cite{Tutusaus:2017ibk}. The CMB does not have direct sensitivity to $\Lambda$ and can only determine $\Omega_k$ and $\Omega_\mathrm{m}$.\footnote{When $\Lambda$ comes to dominate the expansion, this slows down the growth of structure and induces via the `late-ISW effect' a correlation between the CMB and large-scale structure. To detect this at $5\sigma$ would require measuring spectroscopic redshifts for 10 million galaxies uniformly distributed in $0 < z < 1$ over the whole sky \cite{Afshordi:2004kz}.}
It is only via the cosmic sum rule that $\Omega_\Lambda$ is then inferred ( $=1-\Omega_k-\Omega_\mathrm{m}$) to be non-zero. However this sum rule is just the Friedmann-Lema\^{i}tre  equation which is valid only for the FLRW model. We wish to emphasise that it is the widespread (and at times convoluted) use of the FLRW model in cosmological data analysis that has created a strong bias towards the $\Lambda$CDM Universe. 

Whether dark energy is just a manifestation of inhomogeneities is an open question \cite{Buchert:2015iva}. This can impact on analysis of cosmological data in several different ways --- since in the `cosmic web’  that we inhabit, our location (e.g. whether in a low or high density region) is important, as is the direction in which we look (e.g. along filaments or voids), and whether we are Copernican or `tilted’ observers in a bulk flow. The spatial scales being probed by observations are important as the inhomogeneous Universe looks different when averaged on different scales. Ultimately we would like a scale-dependent description of the Universe (akin to the renormalisation group) but this is a long way from the (perturbed) FLRW model which has served so far as the workhorse of cosmology and led us to infer the existence of unphysical dark energy.


\end{document}